\documentclass{kluwer}    

\newdisplay{guess}{Conjecture}

\usepackage{graphicx}
\usepackage{dcolumn}
\usepackage{bm}
\usepackage{latexsym}
\usepackage{amsfonts}
\usepackage{amssymb}
\newcommand{\R }{ \mathop{R^{\mu}_{\ \nu}}^{(4)}  }
\newcommand{\Rs }{ \mathop{R}^{(4)}  }

\begin{document}                                                                                   
\begin{article}

\begin{opening}         
\title{Braneworld Effective Action at Low Energies  }
\author{Sugumi \surname{Kanno}} 
\runningtitle{Brane Effective Action at Low Energies and Radion }
\runningauthor{Sugumi Kanno \& Jiro Soda}
\institute{ Graduate School of Human and Environmental
 Studies, Kyoto University, Kyoto
 606-8501, Japan }
\author{Jiro \surname{Soda}}  
\institute{Department of Fundamental Sciences, FIHS, 
Kyoto University, Kyoto 606-8501, Japan}

\date{September 3, 2002}

\begin{abstract}
The low energy effective theory for the Randall-Sundrum two-brane system 
is investigated with an emphasis on the role of the non-linear radion 
in the brane world.   
It is  shown that the gravity on the brane world is described by 
a quasi-scalar-tensor theory with a specific coupling function.

\end{abstract}


\end{opening}           

\section{Introduction}  

Motivated by the recent development of the superstring theory, 
the brane world scenario has been studied intensively. 
 So far, the works are mostly restricted to the linear theory 
or to homogeneous cosmological models.  
Here, we consider the nonlinear brane gravity
 and derive the effective equations of motion 
for this system using a low energy expansion method 
developed by us~\cite{KS1,KS2}.      

In this paper, we show that the radion 
disentangles the non-locality in the non-conventional Einstein 
equations and leads to the local quasi-scalar-tensor gravity.

\section{Low Energy Expansion}

\subsection{RS1 Model and Basic Equations}  
   
 We  consider an $S_1/Z_2$ orbifold spacetime with the 
 two branes as the fixed points.
In the RS1 model~\cite{RS1}, the two flat 3-branes 
are embedded in the 5-dimensional asymptotically anti-deSitter (AdS) bulk 
with the curvature radius $l$.

For general non-flat branes, we can not keep both of the two branes
straight in the Gaussian normal coordinate system.
Hence, we use the following coordinate system to describe 
the geometry of the brane model;
$
ds^2 = e^{2\phi (x^{\mu})} dy^2 
+ g_{\mu\nu} (y,x^{\mu} ) dx^{\mu} dx^{\nu}  \ .
$
We place the branes at $y=0$ ($A$-brane) and $y=l$ ($B$-brane) 
in this coordinate system. The proper distance between
the two branes
 with fixed $x$ coordinates can be written as
$
   	d(x) = e^{\phi (x)}l \ .
   	\label{eq:distande}
$
Hence, we call $\phi$ the radion.    
We introduce the tensor 
$K_{\mu\nu} = -g_{\mu\nu,y}/2$ for convenience. 
Decompose the extrinsic curvature into the traceless part and the trace part
$
  	e^{-\phi}K_{\mu\nu} 
    		= \Sigma_{\mu\nu} + 1/4~g_{\mu\nu} Q  \ , \quad
    	Q = - e^{-\phi}{\partial \over \partial y}\log \sqrt{-g}    \  .
$
Then, off the brane, we obtain the basic equations;
\begin{eqnarray}
 	& & e^{-\phi} \Sigma^\mu_{\ \nu , y} - Q \Sigma^\mu_{\ \nu} 
    		= -\left[ \R  
            	-\nabla^\mu \nabla_\nu \phi 
               	-\nabla^\mu \phi  \nabla_\nu \phi
       		\right]_{\rm traceless}      \ ,     
       		\label{eq:munu-traceless} \\
	& & {3\over 4} Q^2 - \Sigma^\alpha_{\ \beta} \Sigma^\beta_{\ \alpha} 
    		= \left[ \Rs \right] + {12\over l^2}   \ ,  
    		\label{eq:munu-trace} \\
 	& & e^{-\phi}  Q_{, y} -{1\over 4}Q^2 
       		- \Sigma^{\alpha \beta} \Sigma_{\alpha \beta} 
    		= \nabla^\alpha \nabla_\alpha \phi 
       		+ \nabla^\alpha \phi \nabla_\alpha \phi - {4\over l^2}  \ , 
       		\label{eq:yy}  \\
 	& & \nabla_\lambda \Sigma_{\mu }^{\ \lambda}  
        	- {3\over 4} \nabla_\mu Q = 0   \ ,
        	\label{eq:ymu}
\end{eqnarray}
where the subscript ``traceless" represents the traceless part of the quantity
in the square brackets and $\R$ is the curvature on the brane and $\nabla_\mu $ denotes the
covariant derivative with respect to the metric $g_{\mu\nu}$. 
And the junction conditions are
\begin{eqnarray}
   	\left[ \Sigma^{\mu}_{\nu} 
   		- {3\over 4} \delta^\mu_\nu Q \right] \Bigg|_{y=0}
    		&=& {\kappa^2 \over 2} (-\sigma_A \delta^\mu_\nu 
    		+ T^{A\mu}_{\quad\nu})  \ , 
    		\label{JC:A} \\
    	\left[ \Sigma^{\mu}_{\nu} 
   		- {3\over 4} \delta^\mu_\nu Q \right] \Bigg|_{y=l}
    		&=& -{\kappa^2 \over 2} (-\sigma_B \delta^\mu_\nu + 
    		\tilde{T}^{B\mu}_{\quad\nu})    \ .
    		\label{JC:B}
\end{eqnarray}

\subsection{Low Energy Expansion Scheme}

In this paper, we will consider the low energy 
regime in the sense that the energy density of the matter, $\rho_i$ 
(i=A,B), on a brane is much smaller than the brane tension, i.e.,
$\rho_i /|\sigma_i| \ll 1$. 
In this regime, a simple dimensional analysis, 
$
\rho_i / |\sigma_i| 
	\sim {l \over \kappa^2 |\sigma_i|}\,{\kappa^2\rho_i\over l}
	\sim ({l/L})^2 \ll 1
$	
implies that the curvature on the brane can be neglected compared with the 
extrinsic curvature at low energies.

Our iteration scheme is to write the metric $g_{\mu\nu}$ 
as a sum of local tensors built out of the induced metric on the 
brane, with the number of derivatives increasing with the order
of iteration, that is, 
$ O((l/L)^{2n})$, $n=0,1,2,\cdots$. 
Hence, we seek the metric as a perturbative series 
\begin{eqnarray}
    	&&  g_{\mu\nu} (y,x^\mu ) =
  		a^2 (y,x) \left[ h_{\mu\nu} (x^\mu) 
  		+ g^{(1)}_{\mu\nu} (y,x^\mu)
      		+ g^{(2)}_{\mu\nu} (y, x^\mu ) + \cdots  \right]  \ , 
      		\label{expansion:metric} \\
   	&&  g^{(n)}_{\mu\nu} (y=0 ,x^\mu ) =  0    \ , \quad n=1,2,3,...
   		\label{BC}
\end{eqnarray}
where the factor $a^2 (y,x) $ is extracted because of the reason 
explained later and we put the Dirichlet boundary condition 
$g_{\mu\nu} (y=0, x) =  h_{\mu\nu} (x)$ at the $A$-brane. 

\section{Background Geometry}

As we can ignore the matter at the lowest order, we obtain the vacuum 
brane. Namely, we have an almost flat brane compared with the curvature scale 
of the bulk space-time. The term $\Sigma^{(0)\mu}_{\quad \  \nu} $
is not allowed to exist because of the junction conditions
(\ref{JC:A}) and (\ref{JC:B}). 
Then, it is easy to solve the remaining equations. The result is
$
    	\Sigma^{(0)\mu}_{\quad \  \nu} =0, \ Q^{(0)} = 4/l \ .
    	\label{0:Q}
$
Using the definition 
$
     	K^{(0)}_{\mu\nu}  
     		= - {1\over 2} {\partial \over \partial y} 
        	g^{(0)}_{\mu\nu}  = \frac{1}{l}e^{\phi}g^{(0)}_{\mu\nu}   \   ,
        	\label{0:k}
$
we get the 0-th order metric  as
\begin{equation}
 	ds^2 = e^{2\phi (x)} dy^2 + a^2 (y,x)
        	h_{\mu\nu}(x^\mu ) dx^\mu dx^\nu,\quad 
      	a(y,x) = \exp\left[-e^{\phi (x)y/l}\right] , 
      	\label{0:metric}
\end{equation}
where  the tensor $h_{\mu\nu}$ is  the induced metric on the A-brane.  

 Given the 0-th order solution, junction conditions (\ref{JC:A}) and 
 (\ref{JC:B})  lead to the well known relations 
$
 	\kappa^2 \sigma_A = 6/l \ ,  \kappa^2 \sigma_B 
 		= -6/l \ .
$
Note that $\phi(x)$ and $h_{\mu\nu}(x)$ are arbitrary functions of $x$ at 
the 0-th order.

\section{Holographic Gravity}

\subsection{Bulk Geometry }

The next order solution is obtained by taking into account the 
terms neglected at the 0-th order. 
It is at this order that the effect of matter comes in. 
We obtain the trace part of the extrinsic curvature as
\begin{equation}
   	Q^{(1)} = {l \over a^2} \left[ {1\over 6} R(h) 
      		+ {y e^{\phi} \over l}\left(  \phi^{|\alpha}_{\ |\alpha} 
       		+  \phi^{|\alpha}  \phi_{|\alpha} \right) 
       		- {y^2 e^{2\phi} \over l^2}  \phi^{|\alpha} \phi_{|\alpha} 
       		\right] ,
       		\label{1:Q}
\end{equation}
where the superscript $(1)$ represents the order of the gradient expansion. 
It is convenient to introduce the Ricci
tensor of $h_{\mu\nu}$, denoted by $R^\mu_\nu (h)$.
Hereafter, we omit the argument of the curvature for simplicity. The traceless
part of the extrinsic curvature is 
\begin{eqnarray}
\Sigma^{(1)\mu}_{\quad \nu} &=&  {l \over  a^2 } \left[ 
	{1\over 2}\left( R^\mu_{\ \nu}  
        - {1\over 4} \delta^\mu_\nu R \right)
        + {y e^{\phi} \over l}\left( \phi^{|\mu}_{\ |\nu}  
	-{1\over 4}\delta^\mu_\nu  \phi^{|\alpha}_{\ |\alpha} 
        \right) \right. \nonumber \\ 
       	&& \left. 
       	+\left( {y^2 e^{2\phi} \over l^2} + {y e^{\phi} \over l}\right)
       	\left(  \phi^{|\mu} \phi_{|\nu} -{1\over 4}\delta^\mu_\nu
       	\phi^{|\alpha}  \phi_{|\alpha} \right) 
       	\right]   
        + {\chi^{\mu}_{\ \nu} (x) \over a^4},
        \label{1:sigma}
\end{eqnarray} 
where $\chi^{\mu}_{\nu}$ is an integration constant
with the property $\chi^{\mu}_{\ \mu} =0 $.  And
$\chi^{\mu}_{\nu}$ 
must be transverse  $\chi^{\mu}_{\ \nu|\mu}=0 $ in order to satisfy 
Eq.~(\ref{eq:ymu}). From these results, we can obtain the bulk metric:
\begin{eqnarray}
  	g^{(1)}_{\mu\nu} &=& -{l^2 \over 2 }\left({1\over a^2} -1 \right) 
    		( R_{\mu\nu}  - {1\over 6} h_{\mu\nu} R ) 
    		+ {l^2 \over 2} 
    		\left( {1 \over a^2 }-1 
    		-{2y e^{\phi}\over l}{1\over a^2} \right) \times \nonumber \\
    		&& \quad \left( \phi_{|\mu \nu}  + {1\over 2} h_{\mu\nu}
   		\phi^{|\alpha}  \phi_{|\alpha} \right)   
  		   -{y^2 e^{2\phi} \over a^2} 
  		\left(  \phi_{|\mu}  \phi_{|\nu}-  {1\over 2} h_{\mu\nu}
   		\phi^{|\alpha}  \phi_{|\alpha}  \right) \nonumber \\
    		&& \quad -{l \over 2}\left({1\over a^4} -1 \right)   
    		\chi_{\mu\nu} \ ,
    		\label{1:metric}
\end{eqnarray}
The term of $\chi_{\mu\nu}$ is essentially the Weyl tensor at 
this order. Note that we have obtained the bulk metric in terms of
$ \phi(x)$, $h_{\mu\nu}(x) $ and $ \chi_{\mu\nu}(x) $. 

\subsection{Quasi-Scalar-Tensor Gravity}

 We shall deduce the equations for $ \phi(x),\ h_{\mu\nu}(x) $ and 
 $ \chi_{\mu\nu}(x) $ from  junction conditions. 
 The junction condition at the $A$-brane is written as
\begin{equation}
   	{l\over 2 } G^\mu_{\ \nu} (h ) + \chi^\mu_{\ \nu}
  		= {\kappa^2 \over 2} T^{A\mu}_{\quad \ \nu} \ .
  		\label{1:einstein-A}
\end{equation}
This equation is nothing but the Einstein equations with 
the  generalized dark radiation $\chi_{\mu\nu}$.
It should be noted that $\chi_{\mu\nu}$ is undetermined at this level,
exhibiting the non-local nature of Eq.~(\ref{1:einstein-A}). 

 The junction condition at 
the $B$-brane is given by
\begin{equation}
   	{l\over 2 } G^\mu_{\ \nu} (f ) + {\chi^\mu_{\ \nu} \over \Omega^4}
  		= -{\kappa^2 \over 2} T^{B\mu}_{\quad \ \nu} \ .
  		\label{1:einstein-B}
\end{equation}
where $f_{\mu\nu}$ is the induced metric on the B-brane and 
$\Omega (x) = a(y=l,x)= \exp[-e^{\phi}]$.
Here, the index of $T^{B\mu}_{\quad \ \nu}$ is the energy momentum tensor 
with the index raised by the induced metric on the $B$-brane.
It should be noted that $h_{\mu\nu}$ and $f_{\mu\nu}$ are not independent,
but it is related as $f_{\mu\nu}=\Omega^2h_{\mu\nu}$ at this order. 
Although Eqs.~(\ref{1:einstein-A}) and (\ref{1:einstein-B})
are non-local individually, with undetermined $\chi_{\mu\nu}$,
one can combine both equations to reduce them to local equations
for each brane. This happens to be possible since
$\chi_{\mu\nu}$ appears only algebraically; one can easily eliminate 
$\chi_{\mu\nu}$ from Eqs.~(\ref{1:einstein-A}) and (\ref{1:einstein-B}). 
Defining a new field $\Psi = 1-\Omega^2$,  we find 
\begin{eqnarray}
 	G^\mu_{\ \nu} (h) &=&{\kappa^2 \over l \Psi } T^{A\mu}_{\quad\ \nu}
      	+{\kappa^2 (1-\Psi )^2 \over l\Psi } T^{B\mu}_{\quad\ \nu} 
      	                                 \nonumber \\
      	&+&{ 1 \over \Psi } \left(  \Psi^{|\mu}_{\ |\nu} 
  		-\delta^\mu_\nu  \Psi^{|\alpha}_{\ |\alpha} \right)
  		+{\omega(\Psi ) \over \Psi^2} \left( \Psi^{|\mu}  \Psi_{|\nu}
  		- {1\over 2} \delta^\mu_\nu  \Psi^{|\alpha} \Psi_{|\alpha} 
  		\right),
  		\label{1:STG-1}
\end{eqnarray}
where the coupling function $\omega (\Psi)$ takes the form:
$
  	\omega (\Psi ) = 3\Psi/2(1-\Psi) \ .
$
We can also determine $\chi^{\mu}_{\nu}$ by eliminating $G^{\mu}_{\nu}$ 
from Eqs.~(\ref{1:einstein-A}) and (\ref{1:einstein-B}). Then,
\begin{eqnarray}
	\chi^{\mu}_{\ \nu} &=& -{\kappa^2 (1-\Psi) \over 2 \Psi} 
      		\left( T^{A\mu}_{\quad \nu} + (1-\Psi)T^{B\mu}_{\quad \nu} 
      		\right)
      		                  \nonumber \\ 
      & &    -{l  \over 2 \Psi} \left[ \left(  \Psi^{|\mu}_{\ |\nu} 
  		-\delta^\mu_\nu  \Psi^{|\alpha}_{\ |\alpha} \right)
  		+{\omega(\Psi ) \over \Psi} \left( \Psi^{|\mu}  \Psi_{|\nu}
  		- {1\over 2} \delta^\mu_\nu  \Psi^{|\alpha} \Psi_{|\alpha} 
  		\right) \right]   \ .
  		\label{eq:chi}
\end{eqnarray}
The condition $\chi^\mu_{\ \mu} =0$ gives rise to the field equation for 
$\Psi$:
\begin{equation}
  	\Box \Psi = {\kappa^2 \over l} {T^A + (1-\Psi)T^B \over 2\omega +3}
  		-{1 \over 2\omega +3}{d\omega \over d\Psi} \Psi^{|\mu} 
  		\Psi_{|\mu} \ ,
  		\label{1:STG-2}
\end{equation}
where we have used the explicit form of $\omega (\Psi) $. 
This equation tells us that the trace part of 
the energy momentum tensor determines the radion field and hence the relative
bending of the brane. 
And $\chi_{\mu\nu}$ is determined by the traceless part of the right-hand side 
of Eq.~(\ref{eq:chi}). Remarkably, $\chi_{\mu\nu}$ is now a secondary entity.

The action is derived from the original 5-dimensional action by substituting
the solution of the equation of motion in the bulk. Up to the first order
, we obtain
\begin{eqnarray}
 	S_A  &=& {l \over 2 \kappa^2} \int d^4 x \sqrt{-h} 
     		\left[ \Psi R(h) - {\omega (\Psi ) \over \Psi} 
     		\Psi^{|\alpha} \Psi_{|\alpha} \right]  
     		          \nonumber \\
     	   & &  + \int d^4 x \sqrt{-h} {\cal L}^A 
      		+ \int d^4 x \sqrt{-h} \left(1-\Psi \right)^2 {\cal L}^B  \ .  
      		\label{action:4-dim} 
\end{eqnarray}

Eqs.~(\ref{1:STG-1}) and (\ref{1:STG-2}) are the basic equations 
to be used in cosmological or astrophysical contexts when the
characteristic energy density is less than $|\sigma_i|$.
In contrast to the usual scalar-tensor gravity, this theory  
couples with two kinds of matter, namely, the matters on both  positive 
 and  negative tension branes,  with different 
effective gravitational coupling constants. 
For this reason, we call this theory the quasi-scalar-tensor gravity.    
Thus, the (non-local) Einstein equations (\ref{1:einstein-A}) with 
the generalized dark radiation has transformed into 
the (local) quasi-scalar-tensor gravity (\ref{1:STG-1}) 
with the coupling function $\omega (\Psi) $. 
 
Here, it should be noted that $\chi^{\mu}_{\nu}$ which appeared
in $g^{(1)}_{\mu\nu}$ is a non-local quantity.
 In fact, eliminating $\Psi$ from Eq.~(\ref{eq:chi}) by solving 
Eq.~(\ref{1:STG-2}) yields a non-local expression of $\chi^{\mu}_{\nu}$. 
If we substitute this non-local expression into Eq.~(\ref{1:einstein-A}), 
we will obtain a non-local theory.  Conversely, one can see that
introducing the radion disentangles the non-locality in the 
non-conventional Einstein equations (\ref{1:einstein-A}) and yields the 
quasi-scalar-tensor
gravity given by Eqs.~(\ref{1:STG-1}) and (\ref{1:STG-2}).
 
For completeness, we shall derive the effective action on 
the $B$-brane. Substituting $h_{\mu\nu} =\Omega^{-2} f_{\mu\nu}$ 
into Eq.~(\ref{action:4-dim}) and defining $\Phi = \Omega^{-2} -1$,
we obtain the effective action on the $B$-brane
\begin{eqnarray}
 	S_B  &=& {l \over 2 \kappa^2} \int d^4 x \sqrt{-f} 
     		\left[ \Phi R(f) - {\omega (\Phi ) \over \Phi} 
     		\Phi^{;\alpha} \Phi_{;\alpha} \right]  
     		          \nonumber \\
     	   & &  + \int d^4 x \sqrt{-f} {\cal L}^B
     	   	+ \int d^4 x \sqrt{-f} {\cal L}^A(1+\Phi)^2.  
      		\label{action-B:4-dim} 
\end{eqnarray}
where the coupling function $\omega (\Phi)$ takes the form:
\mbox{$\omega (\Phi ) = -3\Psi/2(1+\Phi)$} and $;$ denote the 
covariant derivative with respect to the metric $f_{\mu\nu}$.

\section{Conclusion}

We have developed a method to deduce the low energy effective theory 
for the two-brane system. 
As a result, we  have shown that the gravity 
on the brane world is described by a quasi-scalar-tensor theory with 
a specific coupling function 
$\omega(\Psi) = 3\Psi /2(1-\Psi ) $ on the positive tension brane
and $\omega(\Phi) = -3\Phi /2(1+\Phi ) $ on the negative tension brane,
where $\Psi$ and $\Phi$ are Brans-Dicke-like scalars on the positive
and negative tension branes, respectively. 

In the process of derivation of the effective equations of motion,  we have 
clarified how the quasi-scalar tensor gravity emerges from  
Einstein's theory with the generalized dark radiation term
described by $\chi_{\mu\nu}$.

\acknowledgements
\theendnotes
We would like to thank M. Sasaki for valuable suggestions. 
This work was supported in part by  Monbukagakusho Grant-in-Aid No.14540258.

\end{article}
\end{document}